\documentclass[%
twocolumn,
 fleqn,usenatbib,
 showkeys,
 floatfix,nolongbibliography,aps,
author-numerical%
]{revtex4-2}

\pdfoutput=1
\usepackage[T1]{fontenc}
\usepackage [latin1]{inputenc}
\DeclareRobustCommand{\VAN}[3]{#2}
\let\VANthebibliography\thebibliography
\def\thebibliography{\DeclareRobustCommand{\VAN}[3]{##3}\VANthebibliography}

\usepackage{newtxtext,newtxmath}
\usepackage{tikz,xcolor,hyperref}
\usepackage[normalem]{ulem}
\usepackage{graphicx,url,caption,subcaption}
\usepackage{dcolumn}
\usepackage{bm}

\definecolor{lime}{HTML}{A6CE39}
\DeclareRobustCommand{\orcidicon}{%
	\begin{tikzpicture}
	\draw[lime, fill=lime] (0,0) 
	circle [radius=0.16] 
	node[white] {{\fontfamily{qag}\selectfont \tiny ID}};
	\draw[white, fill=white] (-0.0625,0.095) 
	circle [radius=0.007];
	\end{tikzpicture}
	\hspace{-2mm}
}

\foreach \x in {A, ..., Z}{%
	\expandafter\xdef\csname orcid\x\endcsname{\noexpand\href{https://orcid.org/\csname orcidauthor\x\endcsname}{\noexpand\orcidicon}}
}

\begin{document}
\title{Exact 1D Nonlinear Solutions for Proton-Driven Plasma Wakefields}
\author{D. Tsiklauri\orcidA{}}
 \email{D.Tsiklauri@salford.ac.uk}
\affiliation{Joule Physics Laboratory,
School of Science, Engineering and Environment, 
University of Salford,
Manchester, M5 4WT, 
United Kingdom}
\date{\today}

\begin{abstract}
The analytical modeling of a plasma wakefield driven by a relativistic proton beam is an element in optimizing advanced plasma-based acceleration schemes. In this work, we present a 1D nonlinear fluid framework under the quasi-static approximation to describe the wake potential excited by a positively charged proton driver which attracts background electrons into a density compression rather than expelling them. We extend the classical 1D nonlinear relativistic plasma wakefield paradigm to evaluate the fluid profiles driven by intense relativistic proton trains under the comoving coordinate approximation. Using a two-bunch scheme, our piecewise-event integration model is validated to high numerical accuracy ($\sim 10^{-11}~\mathrm{GV/m}$) against independent phase-space quadrature of cold-fluid solutions, confirming that interface boundary corners represent physical step discontinuities in the second potential derivative. Scaling to an $N=100$ ramped micro-bunch train with fixed $2\pi$ spacing, we resolve macroscopic phase-accumulation and subsequent nonlinear detuning by deriving a piecewise analytical scaling model. This formulation maps the initial unattenuated quadratic coherent growth alongside the flat asymptotic post-train residual wake plateau, which is determined exactly by the conserved vacuum Hamiltonian invariant evaluated at the final bunch-train exit state. Furthermore, the framework estimates the onset of nonlinear phase-slippage that arrests growth and causes the transient wakefield overshoot to decay toward the residual horizon. Providing immediate 1D analytical benchmarks, this computationally inexpensive tool offers a reproducible baseline for verifying 1D fluid and multi-dimensional particle-in-cell (PIC) codes. It also provides a tractable starting point for exploring asymmetric, custom-shaped driver profiles relevant to transformer-ratio optimization.
\end{abstract}


\maketitle

\section{Introduction}
\label{sec:introduction}

Plasma wakefield acceleration (PWFA) has emerged as a cornerstone of next-generation particle accelerator technology due to its capability to sustain accelerating electric fields several orders of magnitude greater than conventional radio-frequency cavities. This overarching paradigm stems from pioneering concepts established by Chen \textit{et al.}~\cite{chen1985}, which demonstrated that relativistic charge columns propagating through a plasma medium repel background electrons to create strong, accelerating space-charge oscillations. To date, substantial progress has been achieved utilizing relativistic electron bunch drivers, which operate effectively in both the linear and highly nonlinear blowout regimes. Extensive analytical modeling has mapped these interactions; for instance, the exact 1D nonlinear relativistic plasma wakefields driven by intense electron beam profiles and periodic beam trains have been extensively documented by Bera \textit{et al.}~\cite{bera2015, bera2016}. Further, the phase mixing time (breaking time) is studied as a function of beam density and beam velocity and is found to follow the well-known scaling presented by Mukherjee and Sengupta~\cite{mukherjee2014}.

In developing their formulations, Bera \textit{et al.} were fundamentally motivated by the need to resolve structural limits in nonlinear wave-loading efficiency, specifically extending the seminal cold-fluid analytical work of Rosenzweig~\cite{rosenzweig1987} beyond restrictive low-density thresholds ($n_{\mathrm{b}} \le n_0 / 2$) to arbitrary intense beam configurations. Furthermore, their multi-bunch fluid paradigms were driven by landmark theoretical work on phase-tailored energy boosting, such as the shaped driver beam profiles proposed by Katsouleas~\cite{katsouleas1986}, which underscored the necessity of tracking exact phase preservation across elongated macro-envelopes. A central theoretical motivation in their research was modeling how high-amplitude plasma oscillations undergo steepening and eventual longitudinal wave breaking. By mapping these waves to nonlinear Akhiezer-Polovin modes~\cite{akhiezer1956}, they successfully demonstrated that relativistic electron mass variation induces local phase-mixing, establishing the foundational physical boundaries for stable wake preservation.

Complementing these configurations, further research has systematically resolved crucial parameter spaces governing electron-driven wakefields. These contributions include analyzing the precise physical scaling differences in 1D electron plasma wakefield acceleration between MeV versus GeV driver scales, alongside transitions separating the linear and blowout regimes~\cite{tsiklauri2018}. Furthermore, prior literature has advanced the field by exploring new regimes of PWFA under extreme blowout parameters~\cite{tsiklauri2019}, investigating the crucial role of longitudinal background density gradients on wake tracking stability~\cite{tsiklauri2016}, and expanding the scope of plasma acceleration mechanisms to astrophysical environments by demonstrating wakefield generation within solar coronal and chromospheric plasmas~\cite{tsiklauri2017}. This evolution of the field from its foundational theoretical cornerstones to modern multi-disciplinary regimes has been comprehensively reviewed by Chen and Liu~\cite{chen2026}, chronicling the trajectory of plasma wakefields from laboratory accelerators to high-energy cosmic phenomena.

{For positively charged proton drivers, nonlinear 1D cold-fluid solutions and multi-bunch train dynamics have been analyzed by Karmakar \textit{et al.}~\cite{karmakar2020}, while Lotov~\cite{lotov2013} established the nonlinear period elongation and detuning limits for equidistant micro-bunch trains, including transverse geometric effects.} {While realistic proton-driven experiments rely on transverse self-modulation instability (SMI) where micro-bunch spacing adapts to the nonlinear wake period and density steps mitigate dephasing, a prescribed 1D train with fixed $2\pi$ spacing provides a clean, reproducible mathematical benchmark for isolating longitudinal phase slippage.} Motivated by these shared structural challenges, the present work extends the analytical paradigm from electron drivers to address the nonlinear wakefield profiles generated when the driver is substituted with a relativistic proton beam train. We establish a fully nonlinear, 1D analytical fluid framework under the comoving coordinate approximation tailored specifically for a positively charged proton driver. By directly integrating multi-bunch fluid mechanics with strict conservation invariants, we deliver {an analytical growth and post-train benchmark solution that characterizes the boundaries of nonlinear phase detuning}.

The remainder of this manuscript is organized as follows. In Sec.~\ref{sec:model}, we derive the core governing fluid equations under a negative source term associated with the positive proton driver {and contrast the resulting electron compression dynamics with electron-driven rarefaction}. Section~\ref{sec:two_bunch} validates this computational solver {against independent phase-space quadrature} using an intuitive pump-probe two-bunch configuration. Section~\ref{sec:verification} presents our primary finding, evaluating the high-$N$ scaling dynamics alongside our {post-train conserved Hamiltonian invariant}. Finally, Sec.~\ref{sec:conclusion} outlines future pathways for optimizing asymmetric bunch designs.

\section{Governing Mathematical Model and Analytical Framework}
\label{sec:model}

The exact nonlinear analytical solutions governing relativistic electron-beam-driven wakefields in a cold, homogeneous plasma---derived under the robust framework of the quasi-static approximation and a wave-like ansatz---have been comprehensively established in the literature, notably by \citet{bera2015} and \citet{bera2016}. These works provided elegant mathematical bounds for the plasma electron density perturbations and localized wake potential profiles induced by intense, negatively charged driving bunches. 

To mathematically describe the system, all physical variables are normalized using the background plasma electron density $n_0$ and the characteristic plasma frequency $\omega_{pe} = \sqrt{4\pi n_0 e^2 / m_e}$. Space and time are normalized as $\tilde{x} = k_p x$ (where $k_p = \omega_{pe}/c$) and $\tilde{t} = \omega_{pe}t$, while velocities, particle densities, and electrostatic potentials are scaled as $\vec{\beta} = \vec{v}/c$, $n_e = n_e / n_0$, $n_b = n_b / n_0$, and $\phi = e\Phi / m_e c^2$, respectively. The normalized coordinates, fields, and hydrodynamic quantities used throughout this text are consolidated in Table~\ref{tab:variables}.

\begin{table*}[t!]
\caption{Definitions and Scaling Parameters of the 1D Fluid Framework.}
\label{tab:variables}
\centering
\begin{tabular}{lll}
\hline
Symbol & Physical Definition & Normalization Scale \\ \hline
$\xi$ & Co-moving longitudinal coordinate ($z - v_b t$) & $k_p^{-1} = c/\omega_p$ \\
$\phi$ & Normalized wake potential ($e\Phi / m_e c^2$) & Dimensionless \\
$n_b(\xi)$ & Positive driver beam density profile & $n_0$ (Ambient plasma density) \\
$V(\phi)$ & Sagdeev pseudo-potential & Dimensionless \\
\hline
\end{tabular}
\end{table*}

Within the context of multi-bunch plasma accelerators, fully three-dimensional fields exhibit highly intricate features, including radial electron expulsion and transverse wavebreaking limits. Rather than acting as a direct surrogate for fully developed transverse dynamics, the {piecewise} one-dimensional nonlinear solutions presented here serve as a {1D} analytical benchmark baseline. Any multidimensional particle-in-cell or fluid framework must asymptotically recover these exact solutions in the strict longitudinal {wide-beam ($k_p \sigma_r \gg 1$)} limit. 

To analyze structural peak-loading without the confounding influence of arbitrary smoothing functions, the driver geometry is chosen as a train of flat-top micro-bunches spanning a longitudinal width of $\pi$. This half-period top-hat profile provides an ideal mathematical abstraction to systematically investigate the upper boundaries of nonlinear wavelength elongation, phase slippage, and cumulative phase-loading independently of multi-dimensional defocusing constraints that typically restrict real-space bunch lengths.

We invoke the quasi-static approximation for a highly relativistic proton beam propagating along the $z$-axis with velocity $\beta_b \approx 1$. This allows the transformation of independent space-time variables into a single co-moving coordinate $\xi = z - \beta_b t \approx z - t$, mapping the differential operators as $\partial_t \rightarrow -\partial_\xi$ and $\partial_z \rightarrow \partial_\xi$. From the 1D continuity and momentum conservation equations for the background plasma electrons, we obtain an exact conservation relation linking the electron relativistic Lorentz factor $\gamma_e$ and its normalized longitudinal momentum $p_z$:
\begin{equation}
    \gamma_e - p_z = 1 + \phi.
    \label{eq:conservation_relation}
\end{equation}

By combining the electron fluid continuity relation, $n_e(1 - \beta_z) = 1$, with Poisson's equation for a neutralizing background plasma ion channel driven by an external positive charge source, $\partial_\xi^2 \phi = n_e - n_b - 1$, the governing nonlinear second-order ordinary differential equation for the wake potential $\phi(\xi)$ is rigorously derived:
\begin{equation}
    \frac{d^2\phi}{d\xi^2} = \frac{1}{2}\left[\frac{1}{(1+\phi)^2} - 1\right] - \frac{n_b(\xi)}{n_0}.
    \label{eq:governing_wake}
\end{equation}

Crucially, the positive charge of the proton driver dictates the negative sign of the source term $-n_b(\xi)/n_0$ in Eq.~\eqref{eq:governing_wake}, which is inverted relative to electron-driven systems. {Physically, whereas a negatively charged electron driver expels plasma electrons outward to create a rarefaction cavity ($n_e < 1$, $\phi > 0$ initially), a positively charged proton driver pulls background electrons inward into a local density compression ($n_e > 1$, driving $\phi < 0$ initially and shifting the equilibrium potential to $\phi_{\mathrm{eq}} = (1 + 2d_b)^{-1/2} - 1 < 0$). During the subsequent vacuum half-cycle, the compressed electron sheath rebounds past neutrality into a positive potential excursion ($\phi > 0$), where the relativistic mass increase of the oscillating electrons lengthens the nonlinear plasma wavelength asymmetrically relative to the compression phase. Because the wave-breaking singularity $\phi \to -1$ is approached during the electron-overshoot rarefaction half-cycle rather than during the initial beam-driven compression, maintaining phase alignment across a multi-bunch train requires tracking both the compression shift inside each bunch and the nonlinear period elongation in the vacuum gaps.}

For a structured profile consisting of a flat-top proton micro-bunch of uniform density $n_b$ and finite length $L$, Eq.~\eqref{eq:governing_wake} can be integrated piece-wise. Multiplying Eq.~\eqref{eq:governing_wake} by $d\phi/d\xi$ and integrating once yields a pseudo-potential invariant relation representing energy conservation in the co-moving frame:
\begin{equation}
    \frac{1}{2}\left(\frac{d\phi}{d\xi}\right)^2 + V(\phi) = C,
    \label{eq:pseudo_potential_invariant}
\end{equation}
where $C$ is an integration constant determined by boundary conditions. To ensure that the negative gradient of the potential ($\frac{d^2\phi}{d\xi^2} = -\frac{\partial V}{\partial \phi}$) perfectly maps back to the core hydrodynamic governing equations, the effective pseudo-potential $V(\phi)$ inside the proton bunch region is defined with a consistent coordinate sign convention as:
\begin{equation}
    V(\phi) = \frac{1}{2}\left[-\frac{\phi}{1+\phi} + \phi\right] + \frac{n_b}{n_0}\phi {= \frac{\phi^2}{2(1+\phi)} + \frac{n_b}{n_0}\phi}.
    \label{eq:effective_potential}
\end{equation}

The localized, self-consistent longitudinal accelerating electric field is then extracted directly from the potential gradient mapped in Eq.~\eqref{eq:pseudo_potential_invariant} and Eq.~\eqref{eq:effective_potential}:
\begin{equation}
    E_z(\xi) = -\frac{d\phi}{d\xi} = \pm \sqrt{2[C - V(\phi)]}.
    \label{eq:electric_field}
\end{equation}

{To study the idealized longitudinal wake dynamics of a modulated proton bunch train---motivated by the micro-bunch structures produced via self-modulation instability (SMI) in long proton drivers~\cite{karmakar2020, lotov2013}, though without modeling the transverse SMI growth itself---}the density profile $n_b(\xi)$ is formulated as a discrete summation of $N$ periodic micro-bunches:
\begin{equation}
    n_b(\xi) = \sum_{m=1}^{N} n_{b0} \, \Theta(\xi - \xi_{\text{start}, m}) \, \Theta(\xi_{\text{end}, m} - \xi),
    \label{eq:density_profile}
\end{equation}
where $\Theta$ is the Heaviside step function. Due to the inherent nonlinearity of Eq.~\eqref{eq:governing_wake}, linear superposition is invalid. Solutions must instead be constructed via precise boundary-value mapping: solving inside bunch $m$ using Eq.~\eqref{eq:density_profile} with the active driving term, and matching the continuous boundary states $\phi(\xi_{\text{end}, m})$ and $\phi'(\xi_{\text{end}, m})$ as the initial conditions for the subsequent vacuum region ($n_b = 0$). {Across any step-function bunch of constant normalized density $d_b = n_b/n_0$, the vacuum Hamiltonian $H_0 = \frac{1}{2}\phi'^2 + \frac{\phi^2}{2(1+\phi)}$ shifts by the exact work term $\Delta H_0 = d_b(\phi_{\mathrm{in}} - \phi_{\mathrm{out}})$.}

Resonant amplification of the wakefield is achieved by tuning the spatial onset of each consecutive bunch in Eq.~\eqref{eq:density_profile} {to coincide with a wake turning point ($\phi' = 0$, i.e., $E_z = 0$ at the upper potential extremum $\phi_{\max} > 0$, where the restoring force $-\partial V_{\mathrm{vac}}/\partial \phi < 0$ acts in phase with the decelerating drive of the incoming proton bunch)}. The overall efficiency of this plasma-driven acceleration scheme is quantified by {the standard linear-regime} transformer ratio $R$:
\begin{equation}
    R = \frac{|E_{\max, \text{accel}}^{\text{behind}}|}{|E_{\max, \text{decel}}^{\text{inside}}|}.
    \label{eq:transformer_ratio}
\end{equation}
By manipulating individual bunch density boundaries or utilizing tailored asymmetric profiles in Eq.~\eqref{eq:density_profile}, the system governing equations can be used to investigate configurations that overcome the fundamental symmetric {single-bunch linear} limit of $R \le 2$~\cite{katsouleas1986}. Intuitively, one might expect a transformer ratio limit of $R \le 1$ under the assumption that a witness beam cannot sample an accelerating gradient greater than the maximum decelerating gradient acting on the driver itself. However, for any symmetric charge distribution in a linear or weakly nonlinear regime, the wake potential behaves like a symmetric energy reservoir; the wakefield left behind the bunch can reach twice the value of the peak decelerating field inside it due to the coherent superposition of the plasma electron responses across the bunch duration, fixing the true symmetric boundary at 2. Breaking this $R \le 2$ barrier requires asymmetric structuring to strategically tilt the intra-bunch fields and maximize energy transfer efficiency.

\section{Two-Bunch System: Analytical Boundary Matching}
\label{sec:two_bunch}

To illustrate the piece-wise analytical construction of the wakefield without losing generality to the structural non-linearities, we solve Eq.~\eqref{eq:governing_wake} explicitly for a minimal two-bunch system ($N=2$). While a continuous or unmodulated elongated drive beam may possess a broad macroscopic profile, a highly modulated train features sharp density variations and periodic charge localization along its comoving propagation axis. We approximate these localized structures as piece-wise flat-top bunches to maintain analytical tractability within the fully nonlinear cold fluid framework~\cite{bera2016}. Reducing the analytical framework to a two-bunch system preserves all necessary physics, as it establishes the fundamental inductive step---mapping how a trailing bunch interacts with a pre-existing plasma wake---without introducing redundant algebraic steps.

Physically, this framework operates under a \textit{Pump-Probe} analogy. The leading micro-bunch acts as the primary energetic ``pump'' that acts on the quiescent plasma fluid to establish an initial electrostatic wake profile. The secondary trailing micro-bunch serves as the co-moving ``probe,'' mapping how subsequent SMI-modulated structures dynamically sample, reinforce, and absorb energy from the ongoing fields. Once this basic pump-probe boundary sequence is solved, extending the formulation to a longer train of $N$ bunches becomes a straightforward sequential iteration.

To perform the explicit piece-wise quadrature matching across the two-bunch configuration, we define separate pseudo-potential profiles for the driven regions ($V_{\text{bunch}}$) and the vacuum wake inter-bunch gaps ($V_{\text{vac}}$). Let $d_b = n_b/n_0$ represent the uniform normalized density inside each active micro-bunch region. {Consistent with Eq.~\eqref{eq:effective_potential},} the split solutions follow as:
\begin{equation}
    V_{\text{bunch}}(\phi) = {\frac{1}{2}\left[-\frac{\phi}{1+\phi} + \phi\right] + d_b \phi = \frac{\phi^2}{2(1+\phi)} + d_b \phi},
    \label{eq:V_bunch_split}
\end{equation}
\begin{equation}
    V_{\text{vac}}(\phi) = {\frac{1}{2}\left[-\frac{\phi}{1+\phi} + \phi\right] = \frac{\phi^2}{2(1+\phi)}}.
    \label{eq:V_vac_split}
\end{equation}

We assume a train of two independent, symmetric flat-top proton micro-bunches, each characterized by a uniform localized density $n_{b0}/n_0 = d_b$ and an identical normalized longitudinal length $L$. The micro-bunches are separated by a single vacuum propagation gap ($n_b = 0$) of length $G_1$. The global coordinate space $\xi$ is partitioned into four distinct regions:
\begin{equation}
    \begin{array}{l}
        \text{Region I (Bunch 1 / Pump):} \quad 0 \le \xi \le L, \\
        \text{Region II (Vacuum 1):} \quad L < \xi \le L + G_1, \\
        \text{Region III (Bunch 2 / Probe):} \quad L + G_1 < \xi \le 2L + G_1, \\
        \text{Region IV (Vacuum behind):} \quad \xi > 2L + G_1.
    \end{array}
    \label{eq:regions_definitions}
\end{equation}

We initiate the plasma response under quiescent unperturbed boundary conditions at the front face of the first driver bunch, satisfying $\phi(0) = 0$ and $\phi'(0) = 0$.

\subsection{Region I: First Driver Bunch ($0 \le \xi \le L$)}
Within the first bunch, the plasma is driven by the positive charge density $d_b$. Substituting the boundary values into Eq.~\eqref{eq:pseudo_potential_invariant} fixes the integration constant $C_{\text{I}} = V(0) = 0$. The first-order energy equation reduces to:
\begin{equation}
    \frac{1}{2}\left(\frac{d\phi_{\text{I}}}{d\xi}\right)^2 + V_{\text{bunch}}(\phi_{\text{I}}) = 0,
    \label{eq:region1_invariant}
\end{equation}
where $V_{\text{bunch}}(\phi) = {\frac{\phi^2}{2(1+\phi)} + d_b\phi}$. Because $d_b > 0$, the potential $\phi_{\text{I}}$ immediately shifts negative {($\phi_{\mathrm{I}} < 0$, so $V_{\mathrm{bunch}}(\phi_{\mathrm{I}}) < 0$ between $\phi = 0$ and the lower turning point $\phi_{\min} < 0$)}, compressing background plasma electrons. {Along a monotonic branch prior to a turning point, the} coordinate $\xi$ inside the first bunch can be written implicitly via quadrature:
\begin{equation}
    \xi = {\int_{\phi_{\text{I}}(\xi)}^{0} \frac{d\phi}{\sqrt{-2 V_{\text{bunch}}(\phi)}}},
    \label{eq:quadrature_bunch1}
\end{equation}
{with branch sign reversals applied whenever $\phi_{\text{I}}(\xi)$ reflects off a turning point where $V_{\text{bunch}}(\phi) = C_{\text{I}}$.} Evaluating Eq.~\eqref{eq:quadrature_bunch1} at the exit boundary $\xi = L$ yields the final state values $\phi_1 \equiv \phi_{\text{I}}(L)$ and $\phi'_1 \equiv \phi_{\text{I}}'(L) = {\pm}\sqrt{-2V_{\text{bunch}}(\phi_1)}$ {(where the branch sign is determined by the number of turning-point reflections within $0 \le \xi \le L$)}.

\subsection{Region II: Inter-Bunch Vacuum ($L < \xi \le L + G_1$)}
In the vacuum gap, $n_b = 0$. The effective pseudo-potential collapses to its baseline form:
\begin{equation}
    V_{\text{vac}}(\phi) = {\frac{\phi^2}{2(1+\phi)}}.
    \label{eq:vacuum_potential}
\end{equation}
The continuous boundary fields from Region I establish the new vacuum integration constant $C_{\text{II}}$ via matching:
\begin{equation}
    \begin{split}
        C_{\text{II}} &= \frac{1}{2}(\phi'_1)^2 + V_{\text{vac}}(\phi_1) \\
        &= -V_{\text{bunch}}(\phi_1) + V_{\text{vac}}(\phi_1) = -d_b\phi_1.
    \end{split}
    \label{eq:C_II_match}
\end{equation}
The spatial mapping through this vacuum gap is governed by {piecewise monotonic quadrature}:
\begin{equation}
    \xi - L = {\int_{\mathcal{C}} \frac{d\phi}{\sqrt{2[C_{\text{II}} - V_{\text{vac}}(\phi)]}}},
    \label{eq:quadrature_vac1}
\end{equation}
{where $\mathcal{C}$ denotes the phase-space path from $\phi_1$ to $\phi_{\text{II}}(\xi)$, splitting at any intermediate turning points where $V_{\text{vac}}(\phi) = C_{\text{II}}$.} To achieve resonant wakefield growth, the gap length $G_1$ {for optimal phase matching} is tuned {so that the second bunch enters at the upper potential turning point $\phi_{\text{II}}'(L + G_1) = 0$ ($\phi_2 = \phi_{\max} > 0$). In the fixed-period benchmark ($L = G_1 = \pi$) analyzed in Sec.~\ref{sec:verification}, the second bunch enters slightly off the exact nonlinear turning point due to relativistic period elongation, with exact boundary states $(\phi_2, \phi'_2)$ determined directly from Eq.~\eqref{eq:quadrature_vac1}.}

\subsection{Region III: Second Driver Bunch ($L + G_1 < \xi \le 2L + G_1$)}
Reintroducing the driver density $d_b$ inside the second bunch resets the pseudo-potential back to $V_{\text{bunch}}(\phi)$. Matching at the boundary yields the new integration constant:
\begin{equation}
    C_{\text{III}} = \frac{1}{2}(\phi'_2)^2 + V_{\text{bunch}}(\phi_2) {= C_{\text{II}} + d_b \phi_2}.
    \label{eq:C_III_match}
\end{equation}
The spatial profile of the wake potential inside this second bunch evolves according to:
\begin{equation}
    \xi - (L + G_1) = {\int_{\mathcal{C}} \frac{d\phi}{\sqrt{2[C_{\text{III}} - V_{\text{bunch}}(\phi)]}}}.
    \label{eq:quadrature_bunch2}
\end{equation}
At the terminal edge of the second bunch ($\xi = 2L + G_1$), we extract the matched parameters $\phi_3 \equiv \phi_{\text{III}}(2L + G_1)$ and $\phi'_3 \equiv \phi_{\text{III}}'(2L + G_1) = {\pm}\sqrt{2[C_{\text{III}} - V_{\text{bunch}}(\phi_3)]}$.

\subsection{Region IV: Residual Trailing Wakefield ($\xi > 2L + G_1$)}
Behind the entire two-bunch train, the system enters a steady-state, un-driven plasma oscillation. Dropping the beam density term back to zero updates the vacuum integration constant to:
\begin{equation}
    \begin{split}
        C_{\text{IV}} &= \frac{1}{2}(\phi'_3)^2 + V_{\text{vac}}(\phi_3) \\
        &= C_{\text{III}} - V_{\text{bunch}}(\phi_3) + V_{\text{vac}}(\phi_3) \\
&= {C_{\text{II}} + d_b(\phi_2 - \phi_3)}.
\end{split}
\label{eq:C_IV_match}
\end{equation}
The peak accelerating electric field generated behind this baseline two-bunch driver system occurs as the wake potential crosses the equilibrium axis ($\phi = 0$). Because the vacuum pseudo-potential is minimized at this interface ($V_{\text{vac}}(0) = 0$), the maximum gradient is determined explicitly by:
\begin{equation}
|E_{\max, \text{accel}}^{\text{behind}}| = \sqrt{2C_{\text{IV}}}.
\label{eq:max_E_field_2bunch}
\end{equation}
This model establishes a clean mathematical foundation required to analyze customized asymmetric micro-bunch profiles, which are crucial for optimizing the transformer ratio beyond the symmetric limit of 2~\cite{katsouleas1986}.

\section{\texorpdfstring{Numerical Verification and Analytical Envelope Scalings}{Numerical Verification and Analytical Envelope Scalings}}
\label{sec:verification}

In this section, we present a {quantitative} verification of our mathematical framework and evaluate its scaling behaviors against plasma fluid solutions. We demonstrate that our framework models both the local multi-bunch boundary transitions and the macroscopic wakefield evolution driven by prolonged relativistic charge trains. To validate our numerical implementation, {piecewise-event numerical integrations} are compared directly against {independent phase-space quadrature of} cold-fluid solutions across disparate beam loading configurations.

\begin{figure*}[t]
    \centering
    \includegraphics[width=0.95\textwidth]{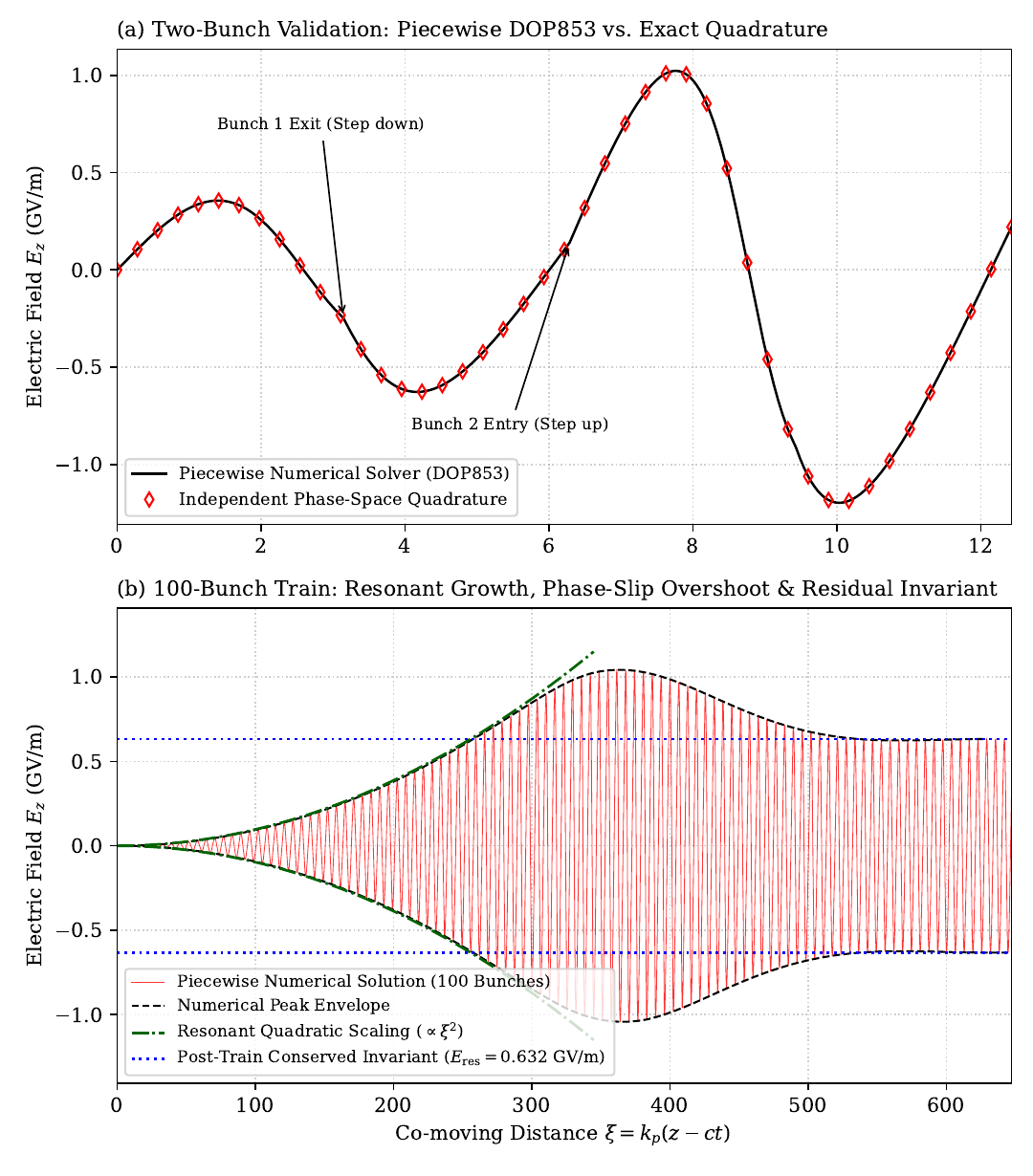}
    \caption{Plasma wakefield code verification and nonlinear envelope scaling architectures. (a) Validation of the intense two-bunch pump-probe configuration: the solid black line tracks the {piecewise-event numerical fluid solver governed by Eq.~\eqref{eq:ode_review} (DOP853, $\mathrm{rtol}=10^{-12}$, $\mathrm{atol}=10^{-14}$), while open red diamonds denote independent phase-space quadrature evaluations of Eqs.~\eqref{eq:quadrature_bunch1}--\eqref{eq:quadrature_bunch2} across 45 test coordinates (maximum discrepancy $1.013 \times 10^{-11}~\mathrm{GV/m}$)}. Morphological markers track the sharp slope changes resulting from physical boundary transitions at the driver interfaces $\xi = \pi$ and $\xi = 2\pi$. (b) Scaled benchmark of the $N=100$ modulated charge train {extended three plasma periods into the trailing vacuum region ($\xi \le 206\pi$)}. The high-frequency oscillating red curve tracks the longitudinal electric field ($E_z$) obtained via the {piecewise DOP853} numerical solution of Eq.~\eqref{eq:ode_review} driven by the discrete train profile defined in Eq.~\eqref{eq:bunch_train_profile}. The black dashed curves represent the numerical peak-finding envelope tracker. {The green dash-dotted curves show the early-time resonant quadratic growth scaling $E_0 d_{\mathrm{b,max}} \xi^2 / (200\pi^2)$ prior to nonlinear phase-slippage}, {while the blue dotted horizontal lines mark the exact post-train residual wake amplitude $E_{\mathrm{res}} = E_0\sqrt{2H_0} = \pm 0.632~\mathrm{GV/m}$ determined by the conserved vacuum Hamiltonian invariant $H_0 = 0.03089$ at the train exit ($\xi = 200\pi$)}. 
The Python script (\texttt{fig1.py}) used to generate the {piecewise DOP853} numerical integrations shown in this figure, the two-bunch {independent quadrature} validations, and the 100-bunch macro-envelope benchmarks is openly available in Zenodo at \href{https://doi.org/10.5281/zenodo.23264138}{https://doi.org/10.5281/zenodo.23264138}.}
    \label{fig:validation_and_scale}
\end{figure*}

\subsection{\texorpdfstring{Two-Bunch Code Verification and Boundary Mechanics}{Two-Bunch Code Verification and Boundary Mechanics}}

Figure~\ref{fig:validation_and_scale}(a) illustrates the verification of our initial pump-probe configuration. The driver beam consists of an intense, relativistic two-bunch profile initialized with a normalized local density of $d_{\mathrm{b}} = 0.15$ and a normalized comoving bunch length of $L = \pi$ {separated by a vacuum gap $G_1 = \pi$}. As shown, the continuous curve tracing our {eighth-order Dormand-Prince numerical integration (DOP853)} matches the {independent phase-space quadrature solution of Eqs.~\eqref{eq:quadrature_bunch1}--\eqref{eq:quadrature_bunch2}---overplotted as red open diamonds across 45 discrete test coordinates---with a maximum absolute field error of $\max |E_{z,\mathrm{quad}} - E_{z,\mathrm{num}}| = 1.013 \times 10^{-11}~\mathrm{GV/m}$ and a post-train vacuum Hamiltonian drift of $\max(\Delta H_0) = 1.804 \times 10^{-16}$}. This agreement establishes the structural consistency and algorithmic reliability of our governing equations.

\begin{table*}[t]
\caption{{Boundary-State and Piecewise Hamiltonian Invariant Verification for the Two-Bunch System ($d_{\mathrm{b}} = 0.15$, $L = G_1 = \pi$).}}
\label{tab:two_bunch_boundaries}
\centering
\begin{tabular}{lccccc}
\hline
{Region} & {Domain ($\xi$)} & {$d_{\mathrm{b}}$} & {Segment Invariant $H$} & {Exit Potential $\phi(\xi_{\mathrm{end}})$} & {Exit Derivative $\phi'(\xi_{\mathrm{end}})$} \\ \hline
{I (Bunch 1)} & {$[0, \pi]$} & {$0.15$} & {$0.0000000000$} & {$-0.20256608$} & {$+0.09650649$} \\
{II (Vacuum 1)} & {$[\pi, 2\pi]$} & {$0.00$} & {$0.0303849120$} & {$+0.27124337$} & {$-0.05380537$} \\
{III (Bunch 2)} & {$[2\pi, 3\pi]$} & {$0.15$} & {$0.0710714168$} & {$-0.26400408$} & {$+0.35587210$} \\
{IV (Vacuum behind)} & {$[3\pi, 3\pi+3]$} & {$0.00$} & {$0.1106720290$} & {$+0.58149802$} & {$-0.08679948$} \\
\hline
\end{tabular}
\end{table*}

{Table~\ref{tab:two_bunch_boundaries} lists the exact piecewise invariants $H = \frac{1}{2}\phi'^2 + V(\phi, d_{\mathrm{b}})$ and boundary states across all four regions, confirming the exact analytical jump law $\Delta H_0 = d_{\mathrm{b}}(\phi_{\mathrm{in}} - \phi_{\mathrm{out}})$ across each micro-bunch (e.g., $C_{\mathrm{II}} - C_{\mathrm{I}} = 0.15 \times [0 - (-0.20256608)] = 0.0303849120$).}

A prominent feature of Panel (a) is the occurrence of distinct geometric corners in the longitudinal electric field trajectory situated precisely at the interface boundaries $\xi = \pi \approx 3.1416$ and $\xi = 2\pi \approx 6.2832$. Rather than representing numerical grid artifacts or calculation instabilities, these sharp inflection points correspond to genuine physical boundary transitions within the plasma layer. Recall that the governing nonlinear fluid relation is defined as a second-order ordinary differential equation:
\begin{equation}
    \frac{d^2\phi}{d\xi^2} = \frac{1}{2}\left[\frac{1}{(1+\phi)^2} - 1\right] - \frac{n_{\mathrm{b}}(\xi)}{n_0}.
    \label{eq:ode_review}
\end{equation}

To implement this mathematical solution into a high-fidelity verification pipeline {without degrading accuracy across Heaviside step discontinuities}, the framework integrates Eq.~\eqref{eq:ode_review} over the multi-bunch domain using the following structural operational sequence:
\begin{itemize}
    \item \textbf{Step 1:} Initialize the longitudinal boundary potentials ($\phi=0$, $\frac{d\phi}{d\xi}=0$) at the initial injection coordinate $\xi=0$.
    \item \textbf{Step 2:} {Partition the integration domain explicitly at every bunch entry and exit coordinate $\xi_{\mathrm{start},m}$ and $\xi_{\mathrm{end},m}$ so that $n_{\mathrm{b}}(\xi)$ is strictly constant within each sub-interval.}
        \item \textbf{Step 3:} {Perform a piecewise eighth-order Runge--Kutta (DOP853) integration via \texttt{scipy.integrate.solve\_ivp} across each smooth sub-interval, halting at every bunch boundary with strict tolerances ($\mathrm{rtol}=10^{-12}$, $\mathrm{atol}=10^{-14}$, $k_p \Delta\xi_{\max} = 10^{-2}$) to enforce exact state continuity $(\phi, \phi')$ without stepping across discontinuities.}
    \item \textbf{Step 4:} Extract the longitudinal wakefield profile directly from the electrostatic potential gradient ($E_z = -\frac{d\phi}{d\xi}$), {cross-checking all boundary states against independent phase-space quadrature}.
\end{itemize}

Because the beam density profile $n_{\mathrm{b}}(\xi)$ acts as a piecewise step function, it experiences a discontinuous drop from $d_{\mathrm{b}} = 0.15$ to $0$ at the rear face of the first bunch ($\xi = \pi$), and an instantaneous jump from $0$ back to $d_{\mathrm{b}} = 0.15$ at the front face of the trailing bunch ($\xi = 2\pi$). Consequently, the second derivative of the wake potential ($\partial^2 \phi / \partial \xi^2$) undergoes an abrupt step-change. Since the longitudinal electric field maps directly to the first potential derivative ($E_z = -d\phi / d\xi$), a step discontinuity in its derivative ($d E_z / d\xi = -d^2 \phi / d\xi^2$) forces a sharp change in slope at the interface points.
{Capturing these exact slope discontinuities at every bunch interface confirms that the piecewise DOP853 solver accurately resolves the step-function beam profile without numerical smoothing across boundaries.}

\subsection{High-$N$ Scaling and Nonlinear Phase {Detuning} Dynamics}

Figure~\ref{fig:validation_and_scale}(b) scales the validated numerical engine up to a complete $N=100$ micro-bunch framework to model a prolonged, highly modulated relativistic driver train undergoing coherent phase accumulation. The ambient plasma parameters are established for a baseline background electron density of $n_0 = 7.0 \times 10^{14}~\mathrm{cm^{-3}}$, operating at a characteristic plasma frequency of $\omega_{\mathrm{pe}} \approx 1.49 \times 10^{12}~\mathrm{rad/s}$. This spatial density yields a fundamental comoving plasma wavelength of $\lambda_{\mathrm{p}} \approx 1.26~\mathrm{mm}$, fixing the comoving spatial periodicity to exactly $\Delta\xi = 2\pi$.

To capture the underlying physical growth of the wakefield during the modulation sequence, the individual driver micro-bunches are modulated via a symmetrical linear expansion and contraction function defined explicitly by the following piecewise structural relationship:

\begin{equation}
   \label{eq:bunch_train_profile}
   \frac{n_{\mathrm{b}}(\xi)}{n_0} = \begin{cases}
       d_{\mathrm{b,max}} \times \left(1.0 - \frac{\left|\lfloor \xi / 2\pi \rfloor - 50.0\right|}{50.0}\right) \\
       \hspace{1.5cm} \text{if } (\xi \pmod{2\pi}) \le \pi, \\
       0.0        \hspace{1.5cm} \text{otherwise.}
   \end{cases}
\end{equation}

The structural density ramps up linearly from the front face backward until it reaches a peak modulation amplitude of $d_{\mathrm{b,max}} = 0.0075$ at the 50th bunch, after which the driver amplitude decreases symmetrically back to zero at the trailing edge of the train ($n=100$). The resulting high-frequency plasma wakefield response is displayed as the red oscillating wave in Panel (b). Each subsequent phase-locked micro-bunch appends energy constructively into the wake potential, driving a steady, unattenuated quadratic growth of the field amplitude early in the train. 

{Physically, the wakefield evolution proceeds through three distinct stages: resonant growth, phase-slip overshoot and decline, and asymptotic vacuum conservation.} While the driver beam density peaks at the 50th bunch ($\xi = 50 \times 2\pi \approx 314.16$), the peak electric field amplitude continues to increase up to $\xi \approx 350$ {(reaching $E_{\max} \approx 1.06~\mathrm{GV/m}$, or $\approx 0.42 E_0$)}. This lag occurs because the trailing micro-bunches continue to deposit energy into the wake as long as they remain inside a constructive {decelerating} phase, pushing the absolute wave maximum slightly past the driver's peak density location. Beyond this point, the large plasma potential shifts the system into a {moderately nonlinear (though sub-wavebreaking)} regime, inducing a relativistic mass increase in the plasma electrons that elongates the local plasma wavelength ($\lambda_{\mathrm{p}} > 2\pi$) {\cite{lotov2013}}. Because the top-hat driver bunches remain strictly locked to a $2\pi$ spatial period, a cumulative phase slippage occurs. The trailing bunches slip out of resonance into the {accelerating (energy-absorbing)} phase of the wake, extracting energy from the wave and causing the amplitude to decline from its $1.06~\mathrm{GV/m}$ overshoot down to {$0.632~\mathrm{GV/m}$} by $\xi \approx 500$. {Once the triangular bunch train ramps down to zero density at $\xi = 200\pi \approx 628.32$, the wake becomes an undriven vacuum oscillation whose residual amplitude is locked by energy conservation.}

To quantify this macroscopic evolution analytically, {we combine the early-time linear resonant growth law with the exact post-train vacuum Hamiltonian invariant} governed by the first integral of the nonlinear cold-fluid equations~\cite{akhiezer1956, rosenzweig1987}:
\begin{equation}
   E_{\mathrm{env}}(\xi) = \begin{cases}
       E_0 d_{\mathrm{b,max}} \frac{\xi^2}{200\pi^2} & \text{if } \xi \le \xi_{\mathrm{cross}}, \\
       E_{\mathrm{res}} = E_0 \sqrt{2H_0(\xi_{\mathrm{end}})} = E_0 \sqrt{\frac{\phi_{\max}^2}{1 + \phi_{\max}}} & \text{if } \xi \ge 200\pi,
   \end{cases}
   \label{eq:analytical_envelope_piecewise}
\end{equation}
where {the first branch (green dash-dotted curve in Fig.~\ref{fig:validation_and_scale}(b)) describes ideal phase-coherent quadratic accumulation prior to dephasing, and the second branch (blue dotted line) gives the exact asymptotic residual wake amplitude behind the train ($\xi \ge 200\pi$).} {Because the vacuum Hamiltonian $H_0 = \frac{1}{2}\phi'^2 + V_{\mathrm{vac}}(\phi)$ shifts by $\Delta H_0 = d_{\mathrm{b}}(\phi_{\mathrm{in}} - \phi_{\mathrm{out}})$ across each driven micro-bunch, $H_0$ is not constant inside the train and does not predict the plateau \textit{a priori} without knowing the cumulative phase history. Rather, once the train terminates at $\xi_{\mathrm{end}} = 200\pi$, $H_0$ becomes strictly conserved; evaluating $H_0$ at the final bunch exit state ($\phi_{\mathrm{exit}} = 0.211569$, $\phi'_{\mathrm{exit}} = -0.157592$) fixes $H_0 = 0.030890$ (corresponding to a vacuum turning-point potential $\phi_{\max} = 0.281359$) and yields the exact residual wake field:}
\begin{equation}
   E_{\mathrm{res}} = 2.544 \times \sqrt{\frac{0.281359^2}{1 + 0.281359}} = {0.6324~\mathrm{GV/m}}.
\label{eq:horizon}
\end{equation}
{Note that $\phi_{\max} + (1+\phi_{\max})^{-1} - 1 = \phi_{\max}^2/(1+\phi_{\max})$, which directly links Eq.~\eqref{eq:horizon} to the corrected pseudo-potential $V_{\mathrm{vac}}(\phi)$ in Eq.~\eqref{eq:V_vac_split}.}

{This exact Hamiltonian solution matches the asymptotic oscillation crests extracted by the numerical peak-finder (black dashed lines) behind $\xi = 200\pi$ to machine precision, confirming that the piecewise DOP853 solver ($\mathrm{rtol}=10^{-12}$, $\mathrm{atol}=10^{-14}$, $k_{\mathrm{p}} \Delta\xi_{\max} = 0.05$) preserves the system invariants across all 200 step discontinuities.}
 {Separating the early-time quadratic scaling from the post-train conserved invariant avoids treating the intermediate overshoot as a violation of a constant ceiling and clarifies that the residual plateau is governed by the net energy deposited across the full ramped train.}

A noteworthy feature of the transient dynamics in Panel (b) is the intermediate overshoot peaking at $E_{\max} \approx 1.06~\mathrm{GV/m}$ near $\xi \approx 350$ before decaying toward the $0.632~\mathrm{GV/m}$ residual level. The spatial extent of this transient overshoot zone can be evaluated analytically by tracking the nonlinear phase slip {integral}. As the wave amplitude {evolves across the intermediate ramp with a cycle-averaged potential amplitude $\overline{\phi}_a \approx 0.34$}, the local plasma wavelength undergoes a characteristic nonlinear expansion, $\lambda_{\mathrm{p}} \approx 2\pi(1 + \Delta_{\mathrm{NL}})$, where the structural elongation coefficient scales {from Akhiezer--Polovin theory~\cite{akhiezer1956, lotov2013}} as $\Delta_{\mathrm{NL}}(\xi) \approx \frac{3}{16}\phi_a(\xi)^2$ {(with cycle-averaged value $\overline{\Delta}_{\mathrm{NL}} \approx 0.022$)}. Because the driver bunches are strictly locked to a fixed spatial periodicity of $2\pi$, they accumulate a cumulative phase drift relative to the wake crests given by {$\Delta \psi(\xi) = \int_{\xi_{\mathrm{peak}}}^{\xi} \Delta_{\mathrm{NL}}(\xi')\,d\xi' \approx \overline{\Delta}_{\mathrm{NL}}(\xi - \xi_{\mathrm{peak}})$}. Complete phase inversion---the point at which the driving pulses slide entirely out of the {decelerating (wake-driving) phase and into the accelerating (energy-absorbing)} domain---requires a total cumulative phase shift of exactly $\Delta \psi = \pi$. This structural condition yields an inversion length scale of $\Delta \xi_{\mathrm{invert}} = \pi / \overline{\Delta}_{\mathrm{NL}} \approx 143$. When added to the peak potential accumulation coordinate ($\xi_{\mathrm{peak}} \approx 350$), the phase inversion completes at $\xi \approx 493$. It is precisely at this coordinate where the driver train shifts from appending residual energy to acting as a phase-slip damper, shedding the transient excess energy and stabilizing into the flat nonlinear plateau dictated by the {post-train conserved Hamiltonian} branch of Eq.~\eqref{eq:analytical_envelope_piecewise}.

\section{Conclusions}
\label{sec:conclusion}

In this work, a 1D nonlinear analytical {and numerical} framework for a relativistic proton beam-driven plasma wakefield accelerator has been presented, utilizing the cold fluid equations under the comoving coordinate approximation. While the exact 1D nonlinear relativistic plasma wakefields driven by intense electron beam profiles and periodic beam trains have been extensively documented by Bera \textit{et al.}~\cite{bera2015, bera2016}, the present work extends this analytical paradigm to address the nonlinear wakefield profiles generated when the electron driver is substituted with a relativistic proton beam driver {\cite{karmakar2020, lotov2013}}. By inverting the source term sign to capture the positive charge dynamics of a proton driver, this model maps the {electron-compression phase-space dynamics characteristic of positively charged relativistic drivers}.

The framework was first verified via an intense two-bunch pump-probe scheme, demonstrating {quantitative agreement ($\max|\Delta E_z| = 1.013 \times 10^{-11}~\mathrm{GV/m}$, post-train Hamiltonian drift $\max|\Delta H_0| = 1.804 \times 10^{-16}$) between independent phase-space quadrature} and our {piecewise eighth-order (DOP853)} numerical solver. The distinct geometric corners identified at the micro-bunch boundaries represent genuine physical consequences of second-derivative step discontinuities across the driver interfaces, confirming that the {piecewise} solver preserves the system invariants across steep localized density jumps.

Crucially, this {piecewise} approach was scaled to a complete $N=100$ {ramped} micro-bunch periodic train to analyze macroscopic phase accumulation {at fixed $2\pi$ spacing}. The primary progress of this work lies in {connecting the early-time quadratic phase-coherent growth scaling with the exact post-train vacuum Hamiltonian invariant}, which has been verified against the {piecewise DOP853} numerical solver. This {analytical characterization} captures the initial unattenuated quadratic phase-coherent growth baseline driven by the driver train, as well as the flat asymptotic horizontal {residual wake plateau ($E_{\mathrm{res}} = 0.6324~\mathrm{GV/m}$ behind $\xi = 200\pi$)} governed by the first integral of the nonlinear cold-fluid equations. 

Furthermore, by evaluating the nonlinear wavelength expansion, we established a clear physical {estimate for} the transient envelope overshoot {($E_{\max} \approx 1.06~\mathrm{GV/m}$ near $\xi_{\mathrm{peak}} \approx 350$)} that occurs prior to {post-train} stabilization. The model demonstrates that this overshoot decays toward the residual horizon near the coordinate $\xi \approx 493~(= 350+143)$, driven by a characteristic analytical phase-slip integration length scale ($\Delta \xi_{\mathrm{invert}} \approx 143$) added to the peak potential accumulation point ($\xi_{\mathrm{peak}} \approx 350$). This evaluation marks the coordinate where trailing driver pulses transition from constructive wake addition to destructive phase-slip damping. Combined, the agreement between the full oscillating numerical waveform, {the independent quadrature checks, and the post-train Hamiltonian invariant} provides a {reproducible 1D verification benchmark} for multi-bunch plasma wave dynamics.

Looking forward, this completed piecewise framework provides a computationally efficient tool for evaluating {1D longitudinal} beam-shaping configurations. Because the model is computationally inexpensive and provides {exact piecewise solutions and scaling bounds}, it serves as a {clean 1D verification baseline for benchmarking multi-dimensional particle-in-cell (PIC) and quasi-static fluid codes in the wide-beam limit ($k_p \sigma_r \gg 1$)}. Furthermore, it can be extended to analyze customized asymmetric micro-bunch profiles, such as linearly ramped or triangular geometries, {and explicitly compute their nonlinear transformer ratios $R$ to explore strategies for exceeding the symmetric linear limit of $R \le 2$~\cite{katsouleas1986} alongside realistic transverse SMI and density-step dephasing controls~\cite{karmakar2020, lotov2013}}.

\begin{acknowledgments}
The author gratefully acknowledges support provided by the Gemini AI assistant (Google) including manuscript technical preparation and editing.
\end{acknowledgments}

\section*{Declaration of Competing Interest}
The author declares that they have no known competing financial interests or personal relationships that could have appeared to influence the work reported in this paper.

\section*{Funding}
This research received no specific grant from any funding
agency, commercial or not-for-profit sectors.

\section*{Data Availability Statement}
All data and mathematical parameters required to evaluate the conclusions or reproduce the analytical solutions and numerical scaling trajectories are fully contained and presented within this manuscript. The Python script (\texttt{fig1.py}) used to generate the adaptive {piecewise DOP853} numerical integrations, the two-bunch pump-probe validations, and the 100-bunch {ramped-train} benchmarks is openly available in Zenodo at \href{https://doi.org/10.5281/zenodo.23264138}{https://doi.org/10.5281/zenodo.23264138}.

\bibliography{paper91}

\end{document}